%Paper: hep-th/9207053
%From: SEN%tifrvax.bitnet@CUNYVM.CUNY.EDU
%Date: Wed, 15 Jul 92 12:15:14 MET DST
%Date (revised): Tue, 6 Oct 92 10:06 IST +0530

% THIS FILE IS TO BE PROCESSED USING PHYZZX.TEX

\input phyzzx.tex

\let\refmark=\NPrefmark 
\def\define#1#2\par{\def#1{\Ref#1{#2}\edef#1{\noexpand\refmark{#1}}}}
\def\con#1#2\noc{\let\?=\Ref\let\<=\refmark\let\Ref=\REFS
         \let\refmark=\undefined#1\let\Ref=\REFSCON#2
         \let\Ref=\?\let\refmark=\<\refsend}

\define\RSTRING
A. Sen, preprint TIFR-TH-92-39 (hepth@xxx/9206016) (to appear in Nucl.
Phys. B).

\define\RNARAIN
K. Narain, Phys. Lett. {\bf B169} (1986) 41.

\define\RNSW
K. Narain, H. Sarmadi and E. Witten, Nucl. Phys. {\bf B279} (1987) 369.

\define\RDGHR
A. Dabholkar, G. Gibbons, J. Harvey and F.R. Ruiz, Nucl. Phys. {\bf B340}
(1990) 33;
A. Dabholkar and J. Harvey, Phys. Rev. Lett. {\bf 63} (1989) 719.

\define\RWITTEN
E. Witten, Phys. Lett. {\bf B153} (1985) 243.

\define\RHOST
G. Gibbons and K. Maeda, Nucl. Phys. {\bf B298} (1988) 741;
D. Garfinkle, G. Horowitz and A. Strominger, Phys. Rev. {\bf D43} (1991)
3140;
G. Horowitz and A. Strominger, Nucl. Phys. {\bf B360} (1991) 197.

\define\RDUALITY
C. Montonen and D. Olive, Phys. Lett. {\bf B72} (1977) 117;
H. Osborn, Phys. Lett. {\bf B83} (1979) 321;
A. Font, L. Ibanez, D. Lust and F. Quevedo, Phys. Lett. {\bf B249} (1990)
35;
J. Harvey and J. Liu, Phys. Lett. {\bf B268} (1991) 40 and references
therein.

\define\RKALLOSH
R. Kallosh, preprint SU-ITP-92-1;
R. Kallosh, A. Linde, T. Ortin, A. Peet and A. Van Proeyen, preprint
SU-ITP-92-13.

\define\RCAMP
B. Campbell, M. Duncan, N. Kaloper and K. Olive, Phys. Lett. {\bf B251}
(1990) 34;
B. Campbell, N. Kaloper and K. Olive, Phys. Lett. {\bf B263} (1991) 364;
{\bf B285} (1991) 199.

\define\RHOL
C. Holzhey and F. Wilczek, preprint IASSNS-HEP-91/71.

\define\RSTROM
A. Strominger, Nucl. Phys. {\bf B343} (1990) 167;
M. Duff, Class. Quantum Grav. {\bf 5} (1988) 189.

\define\RIENGO
M. Fabbrichesi, R. Iengo and K. Roland, preprint SISSA/ISAS 52-92-EP.

\define\RGAZU
M. Gaillard and B. Zumino, Nucl. Phys. {\bf B193} (1981) 221.

\define\RIP
J. Horne, G. Horowitz and A. Steif, Phys. Rev. Lett. {\bf 68}
(1992) 568;
M. Rocek and E. Verlinde, preprint IASSNS-HEP-91-68;
P. Horava, preprint EFI-91-57;
A. Giveon and M. Rocek, preprint IASSNS-HEP-91-84;
S. Kar, S. Khastagir and A. Kumar, preprint IP-BBSR-91-51;
J. Panvel, preprint LTH 282.

\define\RODD
S. Ferrara, J. Scherk and B. Zumino, Nucl. Phys. {\bf B121} (1977) 393;
E. Cremmer, J. Scherk and S. Ferrara, Phys. Lett. {\bf B68} (1977) 234;
{\bf B74} (1978) 61;
E. Cremmer and J. Scherk, Nucl. Phys. {\bf B127} (1977) 259;
E. Cremmer and B. Julia, Nucl. Phys.{\bf B159} (1979) 141;
M. De Roo, Nucl. Phys. {\bf B255} (1985) 515; Phys. Lett. {\bf B156}
(1985) 331;
E. Bergshoef, I.G. Koh and E. Sezgin, Phys. Lett. {\bf B155} (1985) 71;
M. De Roo and P. Wagemans, Nucl. Phys. {\bf B262} (1985) 646;
L. Castellani, A. Ceresole, S. Ferrara, R. D'Auria, P. Fre and E. Maina,
Nucl. Phys. {\bf B268} (1986) 317; Phys. Lett. {\bf B161} (1985) 91;
S. Cecotti, S. Ferrara and L. Girardello, Nucl. Phys. {\bf B308} (1988)
436;
M. Duff, Nucl. Phys. {\bf B335} (1990) 610.

\define\RVENEZIANO
G. Veneziano, Phys. Lett. {\bf B265} (1991) 287;
K. Meissner and G. Veneziano, Phys. Lett. {\bf B267} (1991) 33; Mod. Phys.
Lett. {\bf A6} (1991) 3397;
M. Gasperini, J. Maharana and G. Veneziano, Phys. Lett. {\bf B272} (1991)
277;
M. Gasperini and G. Veneziano, preprint CERN-TH-6321-91.

\define\RSEN
A. Sen, Phys. Lett. {\bf B271} (1991) 295; {\bf 274} (1991) 34.

\define\RFAWAD
S.F. Hassan and A. Sen, Nucl. Phys. {\bf B375} (1992) 103.

\define\RROTBLA
A. Sen, Phys. Rev. Lett. {\bf 69} (1992) 1006.

\define\RMAHSCH
J. Maharana and J. Schwarz, preprint CALT-68-1790 (hepth@xxx/9207016).

\define\RTSW
A. Shapere, S. Trivedi and F. Wilczek, Mod. Phys. Lett. {\bf A6}
(1991) 2677.

\define\RSCHER
J. Scherk and J. Schwarz, Nucl. Phys. {\bf B153} (1979) 61.

\def\eps{\epsilon}
\def\p{\partial}
\def\tF{\tilde F}
\def\bl{\bar\lambda}
\def\Z{{\bf Z}}
\def\R{{\bf R}}

{}~\hfill\vbox{\hbox{TIFR-TH-92-41}\hbox{hepth@xxx/9207053}\hbox{IC-92-171}
\hbox{July, 1992}\hbox{Revised version}}\break

\title{ELECTRIC MAGNETIC DUALITY IN STRING THEORY}

\author{Ashoke Sen}

\address{International Center for Theoretical Physics,
P.O. Box 586, Trieste, I-34100, Italy}
\andaddress{Tata Institute of Fundamental Research,
Homi Bhabha Road, Bombay 400005, India\foot{Permanent address}}
\centerline{e-mail address: sen@tifrvax.bitnet}

\abstract

The electric-magnetic duality transformation in four dimensional
heterotic string
theory discussed by Shapere, Trivedi and Wilczek is shown to be an exact
symmetry of the equations of motion of low energy effective field theory
even after including the scalar and the vector fields, arising due to
compactification, in the effective field theory.
Using this duality transformation we construct rotating black hole
solutions in the effective field theory carrying both, electric and
magnetic charges.
The spectrum of extremal magnetically charged black holes
turns out to be similar to that of
electrically charged elementary string excitations.
We also discuss the possibility that the duality symmetry is an exact
symmetry of the full string theory under which electrically charged
elementary string excitations get exchanged with magnetically charged
soliton like solutions.
This proposal might be made concrete following the suggestion of Dabholkar
et. al. that fundamental strings may be regarded as soliton like
classical solutions in the effective field theory.

\endpage

%\sequentialequations

\chapter{Introduction}

In a recent paper, Shapere et. al.\RTSW\ showed that the equations of
motion of the coupled Einstein-Maxwell-axion-dilaton system, that occur
in the low energy effective action in string theory, is invariant under
an electric-magnetic duality transformation\RGAZU\RODD\
that also interchanges the
strong and weak coupling limits of string theory.\foot{Although the
authors of ref.\RTSW\ claim that the duality symmetry is valid for a
restricted class of backgrounds for which $2F_{\mu\rho}\tF_\nu^{~~\rho}
+2F_{\nu\rho}\tF_\mu^{~~\rho} - G_{\mu\nu} F_{\rho\sigma}
\tF^{\rho\sigma}$ vanishes, it is easy to see that this term vanishes
identically in four dimensions, and does not impose any restriction on
the backgrounds.}
In this paper we shall show that this symmetry is valid even when we
include the extra massless scalar and vector fields that arise from the
(toroidal) compactification of ten dimensional heterotic string theory
to four dimensions.

As was discussed in ref.\RTSW, the duality transformation, together with
the invariance under the shift of the axion field, form an SL(2,\R)
group.
One immediate consequence of this symmetry is that one can use this
transformation on known electrically charged classical solutions to
generate solutions carrying both electric and magnetic charges\RTSW.
\foot{The duality symmetry of the
effective field theory in the absence of the axion field has been
used even before for this purpose\RHOST.}
We apply these transformations on the known rotating charged black hole
solution in string theory\RROTBLA\ to construct rotating dyonic
black hole solutions in
string theory.
We also find that when the string coupling constant in equal to unity
(with suitable normalization), the
spectrum of purely magnetically charged extremal
black holes has a remarkable
similarity with the spectrum of electrically charged elementary string
excitations.

Next we investigate the possibility that the duality symmetry is an
exact symmetry of string theory under which the electrically charged
elementary string excitations get exchanged with magnetically charged
solitons in the theory\RDUALITY\RSTROM.
Since instanton corrections break the symmetry involving translation of
the axion field to the discrete group \Z, this would imply
that string theory has SL(2,\Z) as its symmetry group.
We show that if this is the case, then points in the configuration space
of string theory related by the SL(2,\Z) transformation must be
identified.
We explore some of the consequences of this identification.

This proposal is made more concrete by following the suggestion of
Dabholkar et. al.\RDGHR\ that strings themselves can be regarded as
(possibly singular) classical solutions of the effective field theory.
In that case, the quantization of the zero modes of this classical
solution should reproduce the full spectrum of the string theory.
Thus if the effective field theory has SL(2,\Z) symmetry, then the full
string theory should also have this symmetry.
In this context we show that in the duality invariant effective field
theory that includes the scalar and vector fields arising due to
compactification, the bosonic zero modes of the four dimensional
string like solution of ref.\RDGHR\ are in one to one correspondance to
the bosonic degrees of freedom of heterotic string moving in four
dimensions.
The fermionic degrees of freedom of the fundamental string, on the other
hand, are expected to arise from the fermionic zero modes
generated by supersymmetry transformation of the
original solution.
Thus in order to establish SL(2,\Z) invariance of the full (effective)
string theory, we need to show the duality invariance of the effective
action even after including the fermionic variables, higher
derivative terms, and quantum corrections in the theory.
We do not address this problem in this paper.

\chapter{Duality Symmetry of Low Energy Effective Action}

The first part of the section will contain a review of the results derived
in ref.\RTSW.
In the second part we shall generalise the results to a
background more general than the one considered in ref.\RTSW\ and show
that duality symmetry holds even for this more general background.
We consider critical heterotic string theory in four dimensions with the
extra six dimensions compactified, and, to start with, consider a
restricted set of background field configurations provided by
the metric
$G_{S\mu\nu}$, the antisymmetric tensor field $B_{\mu\nu}$, the dilaton
field $\Phi$, and a single
$U(1)$ gauge field $A_\mu$. The effective action at low energy is given by,
$$\eqalign{
S=&-\int d^4x\sqrt{-\det G_S} e^{-\Phi}(-R_S
-G_S^{\mu\nu}\p_\mu\Phi\p_\nu\Phi  + {1\over 12} G_S^{\mu\mu'}
G_S^{\nu\nu'} G_S^{\tau\tau'}H_{\mu\nu\tau} H_{\mu'\nu'\tau'}\cr
& +{1\over 8} G_S^{\mu\mu'}G_S^{\nu\nu'} F_{\mu\nu} F_{\mu'\nu'})
}
\eqn\etwoone
$$
where,
$$
F_{\mu\nu} = \p_\mu A_\nu -\p_\nu A_\mu
\eqn\etwothreea
$$
$$
H_{\mu\nu\rho} = (\p_\mu B_{\nu\rho} +{\rm ~cyclic~permutations}) -
(\Omega_3(A))_{\mu\nu\rho}
\eqn\etwotwo
$$
$$
(\Omega_3(A))_{\mu\nu\rho} ={1\over 4} (A_\mu F_{\nu\rho} +{\rm ~cyclic
{}~permutations})
\eqn\etwothree
$$
In writing down the action \etwoone, we have set the string coupling
constant to unity.
We can represent a background for which the string coupling constant is
not unity by having an asymptotic value of $\Phi$ different from zero.
The
subscript $S$ denotes that we are using the $\sigma$-model metric,
which is related to the Einstein metric $G_{\mu\nu}$ through the relation
$$
G_{S\mu\nu} = e^{\Phi} G_{\mu\nu}
\eqn\etwofour
$$
{}From now on we shall express all the quantities in terms of the Einstein
metric instead of the $\sigma$-model metric and all the indices will be
raised or lowered with the Einstein metric.
The equations of motion derived from the action \etwoone\ take the form:
$$
R_{\mu\nu} -{1\over 2} D_\mu \Phi D_\nu\Phi -{1\over 2} G_{\mu\nu} D^\rho
D_\rho\Phi
-{1\over 4} e^{-2\Phi} H_{\mu\rho\tau} H_{\nu}^{~~\rho\tau} -{1\over 4}
e^{-\Phi} F_{\mu\rho}F_{\nu}^{~~\rho} =0
\eqn\etwofive
$$
$$
D_\rho (e^{-2\Phi}H^{\mu\nu\rho})=0
\eqn\etwosix
$$
$$
D_\mu(e^{-\Phi}F^{\mu\nu})+{1\over 2} e^{-2\Phi} H_{\rho\mu}^{~~~~\nu}
F^{\rho\mu} =0
\eqn\etwoseven
$$
$$
D^\mu D_\mu\Phi +{e^{-2\Phi}\over 6} H_{\mu\nu\rho}
H^{\mu\nu\rho} +{e^{-\Phi}\over 8} F_{\mu\nu} F^{\mu\nu} =0
\eqn\etwoeight
$$
where $D_\mu$ denotes covariant derivative.
The Bianchi identity for $H_{\mu\nu\rho}$ is given by,
$$
(\sqrt{-\det G})^{-1}\eps^{\mu\nu\rho\sigma}\p_\mu H_{\nu\rho\sigma}
=-{3\over  4} F_{\mu\nu}\tilde F^{\mu\nu}
\eqn\etwonine
$$
where,
$$
\tF^{\mu\nu} ={1\over 2}(\sqrt{-\det G})^{-1} \eps^{\mu\nu\rho\sigma}
F_{\rho\sigma}
\eqn\etwoten
$$
Using eq.\etwosix\ we can define a scalar field $\Psi$ such that,
$$
H^{\mu\nu\rho} =-(\sqrt{-\det G})^{-1} e^{2\Phi}\eps^{\mu\nu\rho\sigma}
\p_\sigma \Psi
\eqn\etwoeleven
$$
Eq.\etwonine\ then gives,
$$
D^\mu (e^{2\Phi}D_\mu \Psi) ={1\over 8} F_{\mu\nu}\tF^{\mu\nu}
\eqn\etwotwelve
$$
Let us now define a complex field $\lambda$ as,
$$
\lambda = \Psi + ie^{-\Phi} \equiv \lambda_1 + i\lambda_2
\eqn\etwothirteen
$$
Eqs.\etwofive, \etwoeight, \etwotwelve\ and \etwoseven\ then gives,
$$
R_{\mu\nu} ={ \p_\mu\bl \p_\nu\lambda +\p_\nu\bl \p_\mu\lambda \over
4(\lambda_2)^2} + {1\over 4} \lambda_2 F_{\mu\rho}F_{\nu}^{~~\rho}
-{1\over 16}\lambda_2 G_{\mu\nu} F_{\rho\sigma} F^{\rho\sigma}
\eqn\etwofourteen
$$
$$
{D^\mu D_\mu\lambda \over (\lambda_2)^2} + i {D_\mu\lambda D^\mu\lambda
\over (\lambda_2)^3} -{i\over 16} F_{-\mu\nu}F_-^{~~\mu\nu} =0
\eqn\etwofifteen
$$
$$
D_\mu (\lambda F_+^{~~\mu\nu} -\bar\lambda F_-^{~~\mu\nu}) =0
\eqn\etwosixteen
$$
where,
$$
F_{\pm} = F\pm i\tF
\eqn\etwoseventeen
$$
In terms of the fields $F_\pm$, the Bianchi identity for $F_{\mu\nu}$
takes the form:
$$
D_\mu (F_+^{~~\mu\nu} - F_-^{~~\mu\nu}) =0
\eqn\etwoeighteen
$$

We now note that eqs.\etwofourteen-\etwosixteen, and \etwoeighteen\ are
invariant under the following transformations:
$$
\lambda \to \lambda + c
\eqn\etwoeighteena
$$
where $c$ is a real number, and,
$$
\lambda \to -{1\over\lambda}, ~~~~ F_+\to -\lambda F_+, ~~~~ F_-\to -\bl
F_-
\eqn\etwoeighteenb
$$
Invariance of all the equations under \etwoeighteena\ is manifest.
Under \etwoeighteenb, eq.\etwofifteen\ is invariant, eqs.\etwosixteen\ and
\etwoeighteen\ get interchanged, and eq.\etwofourteen\ transforms to
itself plus an extra term, given by,
$$
-{\lambda_1 (\lambda_2)^2\over |\lambda|^2} (2 F_{\mu\rho}
\tF_\nu^{~~\rho} + 2 F_{\nu\rho} \tF_\mu^{~~\rho} -g_{\mu\nu} F_{\rho\tau}
\tF^{\rho\tau})
\eqn\etwonineteen
$$
The term given in eq.\etwonineteen, however, vanishes identically in four
dimensions, showing that \etwoeighteenb\ is a genuine symmetry of the
equations of motion.
The two transformations together generate the full SL(2,\R) group under
which $\lambda\to (a\lambda +b)/(c\lambda +d)$ with $ad-bc=1$, and
$F_+\to -(c\lambda +d) F_+$.

\def\ten{{(10)}}
\def\hG{\hat G}
\def\hB{\hat B}
\def\hA{\hat A}

Let us now consider a more general class of field configurations where
some of
the fields originating due to the dimensional reduction from ten to four
dimensions are present.
More specifically, we shall consider heterotic string theory in ten
dimensions, with six of the directions compactified on a torus.
Let $x^\alpha$ denote the ten coordinates, and $G^\ten_{S\alpha\beta}$,
$B^\ten_{\alpha\beta}$, $A^{\ten I}_\alpha$, and $\Phi^\ten$ denote the
ten dimensional fields ($1\le I\le 16$).
(Note that we restrict our background gauge fields to the $U(1)^{16}$
subgroup of the full non-abelian group.)
The subscript $S$ of $G$ again denotes that this is the metric that appear
directly in the $\sigma$ model.
Let $\mu,\nu$ denote the four dimensional indices ($0\le \mu,\nu\le 3$), and
$m,n$ denote the six dimensional indices ($4\le m,n\le 9$).
We shall restrict to background field configurations which depend only on
the four coordinates $x^\mu$.
We now define\RSCHER\RMAHSCH,
$$
\hG_{mn} = G^\ten_{Smn}, ~~~~ \hB_{mn} = B^\ten_{mn}, ~~~~ \hA^{I}_m =
A^{\ten I}_m
\eqn\etwotwenty
$$
$$
C^m_\mu = \hG^{mn}G^\ten_{Sn\mu}, ~~~~ A^{I}_\mu = A^{\ten I}_\mu
-\hA^{I}_m C^m_\mu, ~~~~ D_{m\mu} = B^\ten_{m\mu} -B_{mn}C^n_\mu
-{1\over 4}\hA^{I}_m A^{I}_\mu
\eqn\etwotwentyone
$$
$$
G_{S\mu\nu}= G^\ten_{S\mu\nu} - G^\ten_{Sm\mu} G^\ten_{Sn\nu} \hG^{mn},
{}~~~~ B_{\mu\nu} =B^\ten_{\mu\nu} - B^\ten_{mn} C^m_\mu C^n_\nu
-{1\over 2}(C^m_\mu D_{m\nu} -C^m_\nu D_{m\mu})
\eqn\etwotwentytwo
$$
$$
\Phi =\Phi^\ten - {1\over 2}\ln\det\hG
\eqn\etwotwentythree
$$
$$
F^{(A) I}_{\mu\nu} =\p_\mu A^{I}_\nu - \p_\nu A^{I}_\mu, ~~~~
F^{(C) m}_{\mu\nu} = \p_\mu C^m_\nu -\p_\nu C^m_\mu, ~~~~
F^{(D)}_{m\mu\nu} = \p_\mu D_{ m\nu} - \p_\nu D_{m\mu}
\eqn\etwotwentyfour
$$
$$
H_{\mu\nu\rho} = [(\p_\mu B_{\nu\rho} +{1\over 2} C^m_\mu
F^{(D)}_{m\nu\rho} +{1\over 2}D_{m\mu}F^{(C)m}_{\nu\rho})
+{\rm~ cyclic~ permutations}] -(\Omega_3(A))_{\mu\nu\rho}
\eqn\etwotwentyfive
$$
$$
F_{\mu\nu} =\pmatrix{ F^{(C)m}_{\mu\nu}\cr F^{(D)}_{m\mu\nu} \cr
-{1\over\sqrt 2}F^{(A)I}_{\mu\nu}\cr}
\eqn\etwotwentysix
$$
In these equations $\hG^{mn}$ denotes components of inverse of the matrix
$\hG_{mn}$.
We also define $28\times 28$ matrices $M$ and $L$ as,
$$
M =\pmatrix{ P & Q & R\cr Q^T & S & U\cr R^T & U^T & V}
\eqn\etwotwentyseven
$$
$$
L=\pmatrix{0 & I_6 & 0\cr I_6 & 0 & 0\cr 0 & 0 & -I_{16}\cr}
\eqn\etwotwentyeight
$$
where $I_n$ denotes $n\times n$ identity matrix, $T$ denotes transpose of
a  matrix, and,
$$\eqalign{
P^{mn} = &\hG^{mn}, ~~~~ Q^m_{~~n} = \hG^{mp} (\hB_{pn}+{1\over 4}\hA^I_p
\hA^I_n), ~~~~ R^{m I} = {1\over\sqrt 2} \hG^{mp}\hA^I_p\cr
S_{mn}= &(\hG_{mp}-\hB_{mp}+{1\over 4}\hA^I_m \hA^I_p)\hG^{pq} (\hG_{qn}
+\hB_{qn} +{1\over 4} \hA^J_q \hA^J_n)\cr
U_m^{~~I}= &{1\over\sqrt 2} (\hG_{mp}-\hB_{mp} +{1\over 4}\hA^J_m \hA^J_p)
\hG^{pq}\hA^I_q, ~~~~ V^{IJ} =\delta^{IJ} +{1\over 2}\hA^I_p \hG^{pq}
\hA^J_q \cr
}
\eqn\etwotwentysevena
$$
Note that these matrices $M$ and $L$ are related to similar matrices
defined in ref.\RFAWAD\ by a similarity transformation.
In terms of these variables, the action of ref.\RFAWAD\ (with $D=10$,
$p=16$)
may be written as\RMAHSCH,
$$\eqalign{
S =& \int d^4 x\sqrt{-\det G_S} e^{-\Phi} \Big(R_S +
G_S^{\mu\nu}\p_\mu\Phi
\p_\nu\Phi -{1\over 12}G_S^{\mu\mu'} G_S^{\nu\nu'} G_S^{\tau\tau'}
H_{\mu\nu\rho} H_{\mu'\nu'\tau'}\cr
& -{1\over 4} G_S^{\mu\mu'} G_S^{\nu\nu'} F^T_{\mu\nu} LML F_{\mu'\nu'}
+{1\over 8}G_S^{\mu\nu} Tr (\p_\mu M L \p_\nu M L)\Big)\cr
}
\eqn\etwotwentynine
$$
Let us now define $G_{\mu\nu}$, $\Psi$ and $\lambda$ through
eqs.\etwofour,
\etwoeleven, and \etwothirteen.
We also define $\tF$ through eq.\etwoten\ and,
$$
F_{\pm\mu\nu} = -MLF_{\mu\nu}\pm i\tF_{\mu\nu}
\eqn\etwoexone
$$
keeping in mind that $F$ now is a 28 dimensional column vector.
Using the relations,
$$
M^T = M, ~~~~ M^T L M = L
\eqn\etwoextra
$$
the equations of motion derived from the action \etwotwentynine\ and
the Bianchi identities can be shown to be invariant under the duality
transformation:
$$
G_{\alpha\beta}\to G_{\alpha\beta}, ~~~~ \lambda \to -{1\over \lambda},
{}~~~~ F_+\to -\lambda F_+, ~~~~ F_-\to -\bar\lambda  F_-
\eqn\etwothirty
$$
Proof of invariance of the equations of motion of the metric, gauge
fields, and the dilaton-axion field follows exactly as before.
The only new feature appears in the proof of the equation of motion of
$G_{mn}$, $B_{mn}$ and $A_m^I$.
Let $\{\phi_i\}$ denote these set of fields.
The dependence of the action on these fields is through the matrix $M$
appearing in the last two terms in the action.
Of these, the last term is manifestly invariant under the duality
transformation, hence the contribution of this term to the equation of
motion $\delta S/\delta\phi_i=0$ is duality invariant.
On the other hand, the contribution to the left hand side of the
equation of motion from the last but one term is given by,
$$
-{\lambda_2\over 4}\sqrt{-\det G} G^{\mu\mu'} G^{\nu\nu'}
F^T_{\mu\nu} L{\delta
M\over\delta  \phi_i} L F_{\mu'\nu'}
\eqn\efirsteqmot
$$
{}From eq.\etwothirty\ we see that under the duality transformation,
$$
F\to -\lambda_1 F -\lambda_2 ML\tF
\eqn\esecondeqmot
$$
Transforming eq.\efirsteqmot\ by this transformation, and using the
equation,
$$
{\delta M\over \delta\phi_i} LM +ML{\delta M\over\delta\phi_i} =0
\eqn\ethirdeqmot
$$
we see that eq.\efirsteqmot\ is invariant under the duality
transformation.

\chapter{Rotating Dyonic Black Holes}

\def\den{(\rho^2 +a^2\cos^2\theta +2m\rho\sinh^2{\alpha\over 2})}

In ref.\RROTBLA\ a solution of the equations of motion derived from the
effective action given in eq.\etwoone\ was found which represents rotating
charged black hole. The solution was given by,
$$\eqalign{
ds^{2} =& -{\rho^2 +a^2\cos^2\theta -2m\rho\over \rho^2
+a^2\cos^2\theta +2m\rho\sinh^2{\alpha\over 2}} dt^2
+{\rho^2
+a^2\cos^2\theta +2m\rho\sinh^2{\alpha\over 2}\over \rho^2 +a^2 -2m\rho}
d\rho^2 \cr
& +(\rho^2
+a^2\cos^2\theta +2m\rho\sinh^2{\alpha\over 2}) d\theta^2
-{4m\rho a\cosh^2{\alpha\over 2}\sin^2\theta\over \rho^2
+a^2\cos^2\theta +2m\rho\sinh^2{\alpha\over 2}} dtd\phi\cr
& +\{(\rho^2+a^2)(\rho^2+a^2\cos^2\theta) +2m\rho a^2\sin^2\theta +4m\rho
(\rho^2+a^2) \sinh^2{\alpha\over 2}+ 4m^2\rho^2\sinh^4{\alpha\over 2}\}\cr
& \times {\sin^2\theta \over \rho^2
+a^2\cos^2\theta +2m\rho\sinh^2{\alpha\over 2}} d\phi^2\cr
}
\eqn\ethreeone
$$
$$
\Phi =-\ln {\rho^2 +a^2\cos^2\theta +2m\rho\sinh^2{\alpha\over 2} \over
\rho^2 +a^2\cos^2\theta}
\eqn\ethreetwo
$$
$$
A_\phi = -{2m\rho a\sinh\alpha\sin^2\theta\over \rho^2 +a^2\cos^2\theta
+2m\rho\sinh^2{\alpha\over 2}}
\eqn\ethreethree
$$
$$
A_t = {2m\rho\sinh\alpha\over \rho^2 +a^2\cos^2\theta
+2m\rho\sinh^2{\alpha\over 2}}
\eqn\ethreefour
$$
$$
B_{t\phi} = {2m\rho a\sinh^2{\alpha\over 2}\sin^2\theta \over \rho^2
+a^2\cos^2\theta +2m\rho\sinh^2{\alpha\over 2}}
\eqn\ethreefive
$$
The other components of $A_\mu$ and $B_{\mu\nu}$ vanish.
(Note that $ds^2$ here denotes the Einstein metric which was called
$ds_E^{\prime 2}$ in ref.\RROTBLA.)
$m$, $a$ and $\alpha$ are three parameters which are related to the mass,
angular momentum, and charge of the black hole.
We shall now perform an SL(2,\R) transformation on this solution to
construct a new solution, and show that this new solution represents a
black hole carrying mass, angular momentum, and electric and magnetic type
charges.
To do this we first need to calculate the various field strengths
associated with the solution given in eqs.\ethreeone-\ethreefive.
They are given by,
$$
F_{\rho\phi} ={2ma\sinh\alpha\sin^2\theta (\rho^2-a^2\cos^2\theta) \over
(\rho^2 +a^2\cos^2\theta +2m\rho\sinh^2{\alpha\over 2})^2}
\eqn\ethreesix
$$
$$
F_{\theta\phi} =- {4m\rho a\sinh\alpha (\rho^2 +a^2
+2m\rho\sinh^2{\alpha\over 2})\sin\theta\cos\theta \over \den^2}
\eqn\ethreeseven
$$
$$
F_{\rho t} = -{2m\sinh\alpha (\rho^2 -a^2\cos^2\theta)\over \den^2}
\eqn\ethreeeight
$$
$$
F_{\theta t} ={4m\rho a^2\sinh\alpha\sin\theta\cos\theta \over\den^2}
\eqn\ethreenine
$$
$$
e^{-2\Phi}\sqrt{-G}H^{\rho t\phi} ={2ma\sinh^2{\alpha\over 2} (\rho^2 -
a^2\cos^2\theta)\sin\theta\over (\rho^2 +a^2\cos^2\theta)^2}
\eqn\ethreeten
$$
$$
e^{-2\Phi}\sqrt{-G} H^{\theta\phi t} ={4m\rho a\sinh^2{\alpha\over
2}\cos\theta \over (\rho^2 +a^2\cos^2\theta)^2}
\eqn\ethreeeleven
$$
{}From eqs.\etwoeleven, \ethreeten\ and \ethreeeleven\ we get,
$$
\Psi={2ma\sinh^2{\alpha\over 2}\cos\theta \over \rho^2 + a^2\cos^2\theta}
\eqn\ethreetwelve
$$
Also, using the definition of $\tF$ given in eq.\etwoten, we get,
$$
\tF_{\rho\phi}={4m\rho a^2\sinh\alpha\sin^2\theta\cos\theta \over \den^2}
\eqn\edualone
$$
$$
\tF_{\theta t} = -{2ma\sinh\alpha(\rho^2- a^2\cos^2\theta)\sin\theta
\over\den^2}
\eqn\edualtwo
$$
$$
\tF_{\theta\phi} = {2m\sinh\alpha\sin\theta (\rho^2 - a^2\cos^2\theta)
(\rho^2 + a^2 + 2m\rho\sinh^2{\alpha\over 2})\over\den^2}
\eqn\edualthree
$$
$$
\tF_{\rho t} =-{4m\rho a\sinh\alpha \cos\theta \over\den^2}
\eqn\edualfour
$$

We can now generate a new solution by performing an SL(2,\R) transformation
on the above solution in the same manner as in ref.\RTSW.
We consider the SL(2,\R) transformation $\lambda\to -(1+c^2)/(\lambda+c)$,
$F_+\to -(\lambda +c)F_+/\sqrt{1 +c^2}$.
The transformed solution is given by,
$$
\lambda ' =-{1+c^2\over \lambda +c}, ~~~~ ds^{\prime 2} = ds^2, ~~~~
F'_{\mu\nu} = - {\Psi+c\over \sqrt{1+c^2}} F_{\mu\nu} +
{e^{-\Phi}\over \sqrt{1+c^2}} \tF_{\mu\nu} \
\eqn\ethreethirteen
$$
We shall not write out the solution in detail, but only give the
asymptotic behavior of the solution in order to identify its electric and
magnetic charges.
The leading components of the electromagnetic field are given by,
$$
F'_{\rho t} \simeq {2mc\sinh\alpha\over \sqrt{1+c^2} \rho^2}, ~~~~
F'_{\theta\phi} \simeq {2m\sinh\alpha\sin\theta\over\sqrt{1+c^2}}
\eqn\ethreefourteen
$$
With appropriate normalization (which sets the coefficient of $F^2$ term
in the action to unity),  the electric and magnetic charges carried
by the solution may be identified to,
$$
Q_{el}={1\over\sqrt 2} {mc\sinh\alpha\over\sqrt{1+c^2}}, ~~~~
Q_{mag} = {1\over\sqrt 2}{m\sinh\alpha\over\sqrt{1+c^2}}
\eqn\ethreeseventeen
$$
Since the metric remains the same, the expressions for the mass $M$ and
angular momentum $J$ of the black hole in terms of the parameters $m$, $a$
and  $\alpha$ remain the same as in ref.\RROTBLA\ and are given by,
$$
M={m\over 2}(1+\cosh\alpha), ~~~~ J={ma\over 2} (1+\cosh\alpha)
\eqn\ethreetwentytwo
$$
As in ref.\RROTBLA, the coordinate singularities of the metric occur on
the surfaces,
$$
\rho^2 -2m\rho +a^2 =0
\eqn\ethreetwentythree
$$
and give rise to the horizons which shield the genuine singularities of
the solution.
The extremal limit is approached as $|a|\to m$ from below, when the
horizon disappears leaving behind naked singularities.
Using eqs.\ethreeseventeen\ and \ethreetwentytwo\ we see that in terms of
the physical parameters the extremal limit corresponds to
$$
M^2\to |J|+{(Q_{el})^2 +(Q_{mag})^2\over 2}
\eqn\ethreetwentyfoura
$$
The expressions for the various fields around a rotating dyonic black hole
was found in ref.\RCAMP\ to linear order in $Q_{el}$ and $Q_{mag}$.

If we start from the action
\etwotwentynine, then we can construct more general dyonic solutions
carrying electric and magnetic charges associated with the gauge fields
$A_\mu^I$, $C_\mu^m$ and $D_{m\mu}$ as follows.
We start from the charge neutral Kerr
solution as in ref.\RROTBLA, and
then perform the $O(6,1)\times O(22,1)$
transformation\RODD\con\RVENEZIANO\RSEN\RIP\noc\RFAWAD\ (as in
ref.\RROTBLA)
that  mixes
the time coordinate with the six left moving and twenty two right moving
coordinates.
This gives rise to a solution carrying electrical type $A_\mu^I$,
$C_\mu^m$ and $D_{m\mu}$ charges.
The explicit construction may be carried out in two stages.
In the first stage, we perform an $O(22)$ transformation on the solution
given by eqs.\ethreeone-\ethreefive.
This generates a solution with 21 extra parameters $n^I$, $p^m$ ($1\le
I\le 16$, $1\le m\le 6$, $\sum_I n^In^I +\sum_m p^m p^m =1$), in terms of
which various components of the gauge fields $A_\mu^I$, $C^m_\mu$ and
$D_{m\mu}$ are given by,
$$
A_\mu^I= n^I A_\mu, ~~~~~ C^m_\mu = -D_{m\mu} = {1\over 2} p^m A_\mu
\eqn\edddone
$$
where $A_\mu$ are given by eqs.\ethreethree\ and \ethreefour.
The metric, the dilaton, and the antisymmetric tensor fields remain
unchanged.

In the next stage, we reexpress the solution as a solution in ten
dimensions using the relations given in eqs.\etwotwenty-\etwotwentythree,
and transform it by a Lorentz boost that mixes the time coordinate with
the six internal coordinates.
The independent parameters may be characterized by the magnitude of the
boost and a five dimensional unit vector in the six dimensional internal
space denoting the direction of the boost.
Since the results are somewhat messy to write down, we shall not write
them down here.
In particular, note that now, besides the gauge fields, the metric and the
axion-dilaton fields, the scalar fields characterizing the matrix $M$ also
acquire non-trivial configuration.
In order to show that the singularities of the resulting four dimensional
metric at $\rho^2 -2m\rho +a^2=0$ are coordinate singularities, we need to
show the existence of explicit gauge and coordinate transformations which
make the field configuration non-singular on this surface.
This is relatively straightforward to show for non-rotating solutions
($a=0$), and we shall restrict ourselves to this case from now on.

The asymptotic behaviour of the transformed solution is easier
to calculate, since the $O(6,1)\times O(22,1)$ transformation laws of
various fields are
simple if we keep terms upto linear order in the fields\RSEN\RFAWAD.
For this we start from the charge neutral Schwarzchild solution and
perform an $O(6,1)\times O(22,1)$ transformation on it directly.
If we define the $6$ and $22$ dimensional
vectors $\vec Q^{(el)}_{L}$ and $\vec Q^{(el)}_{R}$ respectively through the
asymptotic relations:
$$
{1\over 2\sqrt 2} (F^{(C)m}_{\rho t} + F^{(D)}_{m\rho t}) \simeq
{(Q^{(el)}_{L})_m\over r^2}
\eqn\esssone
$$
$$
\pmatrix{{1\over 2\sqrt 2} (F^{(C)m}_{\rho t} - F^{(D)}_{m\rho t})\cr
{1\over 2\sqrt 2} F^{(A)I}_{\rho t}\cr
} \simeq {1\over r^2} \pmatrix{ (Q^{(el)}_R)_m\cr (Q^{(el)}_R)^I\cr }
\eqn\esstwo
$$
then the mass $M$ and the charges $\vec Q_{eL}$
and
$\vec Q_{eR}$ of the
final solution may be expressed in terms of the parameters of the
original solution as,
$$\eqalign{
M =& {1\over 2} m (1 +\cosh\alpha\cosh\beta)\cr
\vec Q^{(el)}_{R} = &{m\over \sqrt 2} \cosh\beta\sinh\alpha ~ \vec u\cr
\vec Q^{(el)}_{L} =& {m\over\sqrt 2} \cosh\alpha\sinh\beta  ~ \vec v\cr
}
\eqn\essthree
$$
where $\vec u$ and $\vec v$ are 22 and 6 dimensional unit vectors
respectively, and $\alpha$ and $\beta$ are two boost angles, characterizing
the $O(22,1)$ and $O(6,1)$ rotation matrices.

The solution carrying magnetic type charges may be obtained from this
solution by the SL(2,\R) transformation discussed above.
Again, the steps involved in this construction are purely algebraic, and
hence we shall not carry out the explicit construction of the solution
here.
The mass and angular momenta of the solution are given by the same
expressions as in eq.\essthree; the electric and magnetic charge vectors
of the transformed solution are related to those defined in eq.\essthree\
by the relations:
$$
\vec Q^{(el)\prime}_{{R\atop L}} ={c\over \sqrt{1+c^2}} \vec
Q^{(el)}_{{R\atop L}};
\vec Q^{(mag)\prime}_{{R\atop L}} ={1\over \sqrt{1+c^2}} \vec
Q^{(el)}_{{R\atop L}}
\eqn\essfour
$$

Let us discuss the purely magnetically charged solutions in some detail,
since these solutions, if present, must be regarded as new states in the
spectrum of string theory.
These solutions are obtained by setting $c=0$.
(For convenience, we shall drop the primes from now on.)
In this case, the extremal limit corresponds to
$m\to 0$ keeping the physical parameters $M$, $\vec Q^{(mag)}_{L}$ and
$\vec Q^{(mag)}_{R}$ fixed.
This, in turn, requires that at least one of the angles $\alpha$ or
$\beta$ approach $\infty$.
By analyzing eqs.\essthree, \essfour\ it is easy to see that the condition
for extremality may be expressed in terms of the physical parameters as,
$$
(M^2 - {(\vec Q^{(mag)}_{R})^2\over 2}) (M^2 - {(\vec
Q^{(mag)}_{L})^2\over 2}) =0
\eqn\essfive
$$

We now note the following amusing coincidence.
The spectrum of states in heterotic string theory which do not have any
oscillator excitations,
is given by (in the normalization convention that we have
adopted),
$$
M^2 \simeq {(\vec Q^{(el)}_{R})^2\over 2} \simeq {(\vec
Q^{(el)}_{L})^2\over 2}
\eqn\ethreetwentysix
$$
for $M>>1$.
(Note that this is also the limit of the black hole mass in which
neglecting the higher
derivative terms in
the effective action is justified ).
The angular momentum carried by such states is zero.
The two relations \essfive\
and \ethreetwentysix\ look identical\foot{This also indicates that the
electrically charged elementary string excitations
become extremal charged black holes.
Relationship between string matter and extremal charged black
holes has also been observed for a more restricted class of solutions in
ref.\RIENGO.} if we restrict to states with $(\vec Q^{(mag)}_L)^2 = (\vec
Q^{(mag)}_R)^2$,   except
for the interchange of $(\vec Q^{(el)}_{R}, \vec Q^{(el)}_{L})$ with
$(\vec Q^{(mag)}_{R},
\vec Q^{(mag)}_{L})$.
$(\vec Q^{(el)}_{R}, \vec Q^{(el)}_{L})$ belongs to an even self dual
Lorentzian lattice\RNARAIN\RNSW, and,
if the magnetic charge is quantised according to the rule $\vec
Q^{(mag)}_{R}.\vec Q^{(el)}_{R} + \vec Q^{(mag)}_{L} .\vec Q^{(el)}_{L}=
integer$, then  $(-\vec
Q^{mag}_R, \vec Q^{(mag)}_L)$ belongs to the same lattice.
This seems to suggest that by including the magnetically charged black
holes as
new elementary particles in the theory\foot{The possibility of treating
black holes as elementary particles has been discussed in ref.\RHOL.
Even if the magnetically charged black holes of the type discussed here
are not of the type that behave as elementary particles according to the
analysis of ref.\RHOL, we must still include them as new states in the
spectrum since they cannot be formed as composites of ordinary
electrically charged matter.}
we may be able to establish the
electric magnetic duality as an exact symmetry of string theory.

In principle the similarity in the spectrum could be destroyed by string
loop corrections, but in this case we shall expect that space-time
supersymmetry will prevent the mass charge relation of extremal black
holes from being modified by radiative corrections\RKALLOSH.
The extremalilty condition for black holes carrying finite angular
momentum do not seem to have such   protection against radiative
corrections, and hence the analysis becomes more complicated in this case.
Also we do not yet have the magnetically charged particles which
are dual to
the elementary string states that have arbitrary oscillator excitations.
In the next section we shall explore the possibility of exact
electric-magnetic duality in string theory in some more detail.

\chapter{Can Duality be an Exact Symmetry of String Theory?}

In this section we shall explore the possibility that duality is an exact
symmetry of string theory under which electrically charged elementary
string excitations get exchanged with magnetically charged
solitons\RDUALITY\RSTROM.
We begin by
discussing the consequences of duality being an exact symmetry,
later we shall come back to the question of how the magnetically
charged states that are required for duality to be an exact
symmetry of the theory might be constructed.
We note the following points:

\pointbegin
The symmetry $\lambda\to \lambda+c$ is broken down to
$\lambda\to \lambda+1$ by instanton
corrections.
Hence the complete symmetry group of the theory can at most be SL(2,\Z) and
not SL(2,\R)\RTSW.

\point
Since the fundamental string is an axionic string\RWITTEN, the axion field
changes by 1 as we go around the fundamental string.
Hence in the configuration space we must identify the points related by
the transformation $T: \lambda\to \lambda+1$.
If the duality transformation $S: \lambda\to -1/\lambda$ together with
appropriate transformation on the other fields is to be a genuine symmetry
of the theory, then the points in the configuration space related by the
transformation $STS^{-1}$ must also be identified.
Since $T$ and $STS^{-1}$ can be shown to generate the full SL(2,\Z) group,
this shows that points in the configuration space related by any SL(2,\Z)
transformation must be idetified.

In order to understand the physical significance of this identification,
we note that the weak coupling perturbation theory
is around a vacuum in which the expectation value
of $Im\lambda$ is large, hence the SL(2,\Z) symmetry acting
on a field configuration around this vacuum will take it to another field
configuration far removed from it (before the identification under
SL(2,\Z) is made).
As a result, perturbation theory around such a vacuum is insensitive to
the identification of fields under the SL(2,\Z) transformation.
As an
analogy, let us consider a field theory of a scalar field $\phi$
with potential $(\phi^2-a^2)^2$, with the identification
$\phi\equiv - \phi$.
The vacuum configuration corresponds to $\phi=a$ which is not invariant
under the \Z$_2$ symmetry, and perturbative
quantization of the theory around this point is totally
insensitive to the
fact that in the configuration space the points $\phi$ and
$-\phi$ are identified.
There could, however, be important non-perturbative effects.
For example, without the identification in the field space, there will be
stable domain walls in the system since there is more than one
vacuum configurations.
On the other hand, with the identification, there is only one vacuum
configuration, and hence there is no stable domain wall.

We should also point out that with this identification, at string
coupling constant equal to unity, the magnetically charged solutions
that we have discussed in the last section are not new states.
Instead, a physical state is to be constructed as a linear superposition
of the electrically charged state and  its duality transform.

The proposal that the SL(2,\Z) invariance
is an exact symmetry of the full string theory can be made more concrete
by following the suggestion of Dabholkar et. al.\RDGHR\ that the
elementary string may be regarded as a classical solution of the
effective field theory, and that the collective excitations
associated with the zero modes of the classical
solution correspond to the dynamical degrees of freedom of the
elementary string.
Quantization of these zero modes would then give rise to the spectrum of
the full string theory.
In that case, SL(2,\Z) invariance of the effective field theory
(after including all the higher derivative terms and quantum corrections)
would automatically imply the
SL(2,\Z) invariance of the full string theory, since,
given any classical string configuration, we can construct its
dual by SL(2,\Z) transformation on the corresponding solution in
the effective field theory.

In order to implement this idea, we need to ensure that we
include enough degrees of freedom in the effective field theory so that
the classical solution describing the string has enough number of
deformations (zero modes) which are in one to one correspondence with
the degrees of freedom of the fundamental string.
At the same time, we need to ensure that this effective field theory is
invariant under the duality transformation, and hence should
not, for example, contain any charged field.
We shall now show that the solution of ref.\RDGHR, regarded as a
solution of the (duality invariant) equations of motion in the
effective field theory
described by the action \etwotwentynine, has all the
degrees of
freedom required to describe the bosonic dynamical degrees of
freedom of the
fundamental string.
As was shown in ref.\RDGHR, the fermionic degrees of freedom come from
the supersymmetry transformation of the original solution.
Thus in order to incorporate these degrees of freedom, we must work with
the full ($N=4$) supersymmetric effective field theory by including also
the fermionic fields in the theory, and prove that this theory is
invariant under duality transformation.
We do not address this problem in this paper.
Nor do we address the question as to whether the SL(2,\Z)
symmetry of the equations of motion survive when we include
higher derivative terms and quantum corrections in the effective action.

In order to see how the bosonic
degrees of freedom of the fundamental
string are related to the zero modes of the classical solution, let us
note that the four bosonic zero modes corrsponding to translation of
the solution in the four dimensional space correspond to the variables
$X^\mu$ of the fundamental string.
The remaining degrees of freedom of the fundamental string are the six
internal bosonic coordinates $X^m$, and the sixteen right moving internal
coordinates $Y^I$.
Alternatively we can regard them as six left moving and
twenty two right moving world sheet currents.
These degrees of freedom may be identified to the deformation of the
original solution of ref.\RDGHR\ by the $O(6,1)\times O(22,1)$
transformation  discussed
in ref.\RFAWAD, which mixes the left and right components of the
time coordinate with the left and right components of the internal
coordinates respectively.
Such deformations are possible since the original solution is
independent of time, as well as internal coordinates.
A particular
example was already
discussed in ref.\RSTRING, where by using an $O(1,1)$ transformation of
the solution of ref.\RDGHR, a charged string solution was
constructed.\foot{
I wish to thank A. Strominger for pointing out the relation between this
deformation and the degrees of freedom of the fundamental string.}
To see how the counting of the degrees of freedom work, note that the
$O(6)\times O(22)$ transformation that mixes the internal coordinates do
not modify the original solution.
Hence the number of independent deformations generated by the $O(6,1)\times
O(22,1)$ transformation is equal to the dimension of the coset space
$(O(6,1)/O(6))\times (O(22,1)/O(22))$, which is equal to $6 +
22$.\foot{Actually in this case there is also a set of deformations
belonging to the coset space $O(6,22)/O(6)\times O(22)$\RNARAIN\ which
correspond to changing the lattice of compactification and putting
constant background $A_m^{(I)}$ and $B_{mn}$ fields
in the theory\RNSW.
%The $O(6,22)$ transformations leave the equations of motion invariant, and
%it is quite likely that the duality transformation discussed in the text
%may need to be accompanied by some O(6,22)
%transformation in order that it is an exact symmetry of string theory.
}
The details of the construction are identical to the
corresponding construction of the general electrically charged black holes
discussed in the last section.
This way we can identify the bosonic dynamical degrees of freedom of the
fundamental string to the bosonic deformations of the classical solution
in the effective field theory.
Construction of the general solution deformed by all the bosonic
zero modes involves purely algebraic manipulations, and will not
be given here.

\chapter{Summary}

In this paper we have shown that the electric-magnetic duality
transformation is an exact symmetry of the equations of motion of the
low energy effective field theory arising in string theory.
Using this symmetry we have constructed rotating black hole solution in
string theory carrying both, electric and magnetic charge.
We have also analyzed the possibility that the duality transformation is
an exact symmetry of string theory, by regarding the fundamental string
as a classical solution in the effective field theory.
We found that in order to make this proposal more concrete, we need to
establish that the full supersymmetric effective field theory is
invariant under duality transformation.
We hope to return to this question in the near future.

We conclude by mentioning that if the duality symmetry indeed turns out
to be an exact symmetry of string theory, it will undoubtedly give us
information about non-perturbative features of string theory, since
the duality transformation interchanges the strong and weak coupling
limits of string theory.

\ack
I wish to thank D. Jatkar for discussions, and A. Strominger for a
comment.
I also wish to thank the Institute for Theoretical Physics at the
University of Heidelberg and the International Centre for Theoretical
Physics at Trieste for hospitality, where part of this work was done.

\refout

\end